\documentclass[preprint,12pt]{elsarticle}
\usepackage{amssymb}
\usepackage{booktabs}
\usepackage{dcolumn}
\usepackage{multirow}
\usepackage{adjustbox}
\usepackage{subfig}
\usepackage{graphicx}
\usepackage{natbib} 
\usepackage{booktabs}
\usepackage{lscape}
\usepackage{float}
\usepackage[obeyspaces,hyphens]{url}
\usepackage{empheq}
\usepackage{fancyhdr}
\usepackage{pdfpages} 
\usepackage{mdwlist}
\usepackage{bm}
\usepackage{blindtext}
\usepackage{array}
\usepackage{multicol}
\usepackage{multirow}
\usepackage{mdwlist}
\usepackage{enumitem}
\usepackage[numbers]{natbib}
\usepackage{graphicx}
\usepackage{caption}    
\usepackage{amsmath}
\usepackage{amsthm}
\usepackage{amsfonts}
\usepackage{mathrsfs}
\usepackage{amssymb}
\usepackage{mdframed}
\usepackage{xcolor}
\usepackage{fourier-orns}
\usepackage{geometry}
\usepackage{parskip}
\usepackage{listings}
\usepackage{setspace}
\usepackage{imakeidx}
\usepackage{makeidx}
\usepackage{setspace} 
\usepackage{etoolbox}
\usepackage[
            pdftex
            ,plainpages=false
            ,pdfpagelabels
            ,pagebackref = false
            ,hyperindex
            ,breaklinks=true 
            ,colorlinks=true 
            ,citecolor=blue 
            ,linkcolor=black
            ,urlcolor=blue
            ,filecolor=blue
            ,bookmarksopen=true
            ]{hyperref}
\usepackage{glossaries}
\addtocontents{toc}{\protect\setcounter{tocdepth}{2}}
\numberwithin{equation}{section}
\biboptions{sort&compress}

\newcommand{\be}{\begin{equation}}
\newcommand{\ee}{\end{equation}}
\newcommand{\beq}{\begin{eqnarray}}
\newcommand{\eeq}{\end{eqnarray}}

\newcommand{\ssty}{\scriptscriptstyle}

\newcommand{\dl}{d_{\ssty L}}

\newcommand{\gaml}{\gamma_{\ssty L}}

\makeatletter
\newcommand{\oz}{\mathbin{\mathpalette\make@circled{z}}}
\newcommand{\odl}{\mathbin{\mathpalette\make@circled{\dl}}}
\newcommand{\ogams}{\mathbin{\mathpalette\make@circled{\gaml^\ast}}}
\newcommand{\make@circled}[2]{%
  \ooalign{$\m@th#1\smallbigcirc{#1}$\cr\hidewidth$\m@th#1#2$\hidewidth\cr}%
}
\newcommand{\smallbigcirc}[1]{%
  \vcenter{\hbox{\scalebox{1.6}{$\m@th#1\bigcirc$}}}%
}
\makeatother
\usepackage{array}

\makeindex 
\journal{}
\begin{document}
\begin{frontmatter}
\title{Experimental Physics Laboratory 2: Calculating the Value of Water 
Density using Metal Rod and Water Container}
\author[1]{Osvaldo L. Santos-Pereira}
\address[1]{Physics Institute, Universidade Federal do Rio de Janeiro, Rio de 
Janeiro, Brazil}
\ead{olsp@if.ufrj.br}
\begin{abstract}
This article presents a detailed analysis of an undergraduate physics 
laboratory experiment designed to determine the density of water using 
fundamental measurement techniques and data analysis methods. The 
experimental setup consists of a precision scale, a graduated container 
filled with water, and a suspended metal rod held by a crank, allowing for 
controlled displacement measurements. The primary objective of this 
experiment is to reinforce essential concepts in experimental physics, 
particularly in deriving physical models that correlate measurable 
quantities, performing precise measurements, and analyzing data using 
regression techniques via ordinary least squares methods for fitting data 
into linear models. This article aims to provide students with a theoretical 
and computational aid to explore the physical interpretations of this 
experiment. I developed a theoretical framework to introduce the fundamental 
concepts of hydrostatics, Newtonian mechanics, and the primary equations used 
in the experiment. I supplied Python code with thorough explanations that 
performs analysis on the experiment.
\end{abstract}
\end{frontmatter}

\newpage
\tableofcontents
\newpage
\section{Introduction}

In physical science, one main objective is to formulate a theoretical 
mathematical model to explain physical phenomena, linking the equations and 
formulas directly to physical quantities that can be measured directly in 
experiments and used as input to make predictions for other physical 
quantities that sometimes cannot be measured directly, but only infer \cite{
feynman_lectures_2011, halliday2010physics, nussenzveig_fisica_basica_vol2}. 
One example is the aim of this work. To discuss and analyze the caveats that 
undergraduate students may face in experimental physics class. The subject of 
this article is the determination of water density using a precision scale, a 
graduated container filled with water, and a suspended metal rod held by a 
crank, which allows for controlled displacement measurements. An experiment 
designed in the Physics Institute of Universidade Federal do Rio de Janeiro (
UFRJ) to be taught in the second experimental physics course in STEM 
undergraduate majors \cite{fisexp22024,fisexp2relatorio1}.

One of the many caveats that students may struggle with in experimental 
physics courses is dealing with simple linear regression between two 
variables. The imposition of a linear model, $y = ax + b$, may be misleading 
since not all physical models are linear. This possible misconception of how 
to transform a nonlinear equation into a linear one can be a limiting 
conceptual gap for students, making the presentation of linearization 
techniques for nonlinear equations crucial for them to learn how to apply 
linear regression to experimental data via ordinary least squares methods. 
The ability to extract meaningful physical parameters from the slope and 
intercept of a fitted linear model is of fundamental importance in 
experimental physics \cite{bevington2003, taylor1997, hill2017, cleveland1993, 
holmes2018, vuolo1992}. 

By analyzing the collected data and fitting a least-squares regression line 
to the mass-volume relationship, students can determine the density of water 
as the slope of the best-fit equation \cite{fisexp22024, fisexp2relatorio1}. 
This article explores students' potential misconceptions and challenges they 
may encounter during this process, offering insights into pedagogical 
strategies that can enhance their understanding of experimental physics and 
data analysis. The results emphasize the importance of integrating 
theoretical modeling, systematic measurement techniques, and statistical data 
analysis to improve students' ability to interpret and extract meaningful 
physical quantities from experimental observations.

A caveat that students may face is the experimental setup and how simple 
physical phenomena can alter the results of experimental measurements. For 
instance, friction forces due to the contact of the metal rod with the 
container's surface may change the mass readings on the scale. Or even the 
main differences between selecting which model to use, a mass as a function 
of volume $M \times V$, or the volume as a function of the mass $V \times M$ 
model. The inferred error in the experimental data differs due to the 
experimental setup and error propagation, so one model is not necessarily 
better. Some students might face difficulties with error propagation 
techniques.

Hydrostatics is a branch of fluid mechanics that studies the equilibrium of 
fluids at rest and the forces exerted by or upon them. The fundamental 
principle governing hydrostatics is Pascal's law, which states that a change 
in pressure applied to an enclosed incompressible fluid is transmitted 
undiminished throughout the fluid 
\cite{halliday2010physics, nussenzveig_fisica_basica_vol2, white_fluid_2016}. 

Another crucial concept is Archimedes' principle, which states that a body 
submerged in a fluid experiences an upward buoyant force equal to the weight 
of the displaced fluid 
\cite{halliday2010physics, nussenzveig_fisica_basica_vol2, white_fluid_2016}. 
These physical principles are widely applied in 
engineering, geophysics, and biological systems, forming the theoretical 
foundation for determining fluid densities experimentally.

In this experiment, an undergraduate-level physics setup is used to determine 
the density of water through buoyancy measurements. The setup consists of a 
submerged cylindrical object connected to a spring system, enabling precise 
control over volume displacement. By analyzing the equilibrium conditions 
before and after submersion, the density of water can be inferred using force 
balance equations. The experiment demonstrates the practical application of 
hydrostatic principles and provides students with hands-on experience in 
fluid mechanics experimentation in the laboratory. For a thorough 
introduction to the development of this experiment, see 
Ref.\,\cite{fisexp22024}, and for the documentation template, see 
Ref.\,\cite{fisexp2relatorio1}.

This work is organized as follows: Section 1 presents some methodological 
references on experimental physics laboratory courses and caveats that 
undergraduate students may face during their formative years. Section two 
discusses and explains the experimental setup, presenting the key physical 
variables. The third section summarizes the main pitfalls and caveats 
encountered by the students during the experimental procedure. The fourth 
section presents the theoretical framework and the derivation of the main 
equations used to model the physical phenomena analyzed in this experiment. 
Section 5 introduces fundamental concepts of statistical tools, including 
linear regression, ordinary least squares methods, and error propagation 
methods. The sixth section presents the results and data analysis from the 
experiment, as well as the approach the student should take to investigate 
the physical phenomenon in this experiment. Section seven presents a 
pedagogical discussion on the difficulties faced by the students during the 
experimentation and the writing of the report. Section eight presents a 
Python class guide on using it to create the data analysis needed for the 
experiment. Lastly, the conclusion of this work is presented. The appendix 
presents the derivation for the slope and the intercept for the Ordinary 
Least Squares, as well as the errors for each of those estimators.

\section{Possible caveats faced by students}

Studying experimental physics at the undergraduate level involves a series of 
pedagogical and practical caveats that affect learning outcomes. Holmes and 
Wieman \cite{holmes2018} demonstrate that traditional “cookbook” labs 
contribute little to conceptual understanding, as students tend to follow 
instructions mechanically without engaging in genuine problem-solving. 
Similarly, Erinosho \cite{erinosho2013} reveals that difficulties in 
conceptual comprehension—such as the abstract nature of physics and its 
mathematical rigor—start early in education and persist into higher 
education, particularly in experimental contexts.

A recurring issue is students' struggle with measurement uncertainty. Pessoa 
et al. \cite{pessoa2025} and Geschwind et al. \cite{geschwind2024} 
demonstrate that even after completing several laboratory courses, many 
students still struggle to understand uncertainty propagation and lack 
confidence in comparing results within error margins. Mossmann et al. 
\cite{mossmann2002} reinforce this by pointing out that, despite technological 
aids such as automated data acquisition, students often encounter difficulties 
when interpreting data involving friction and measurement errors.

Another key problem lies in the conceptualization of data itself. Buffler et 
al. \cite{buffler2001} introduce the notion of “point” versus “set” 
paradigms, explaining that novices often fail to consider variability in 
measurements, instead treating single data points as definitive. In Brazilian 
engineering labs, Parreira and Dickman \cite{parreira2020} observe a 
misalignment between students and instructors. While students perceive labs 
as mere reinforcements of theory, instructors seek to develop critical and 
experimental thinking.

Technological interventions, such as educational software or simulations, 
present both benefits and risks, as noted in works by Silva et al. 
\cite{silva2002} and Magalh\~aes et al. \cite{magalhaes2002}. They emphasize 
the value of computational tools to support visualization and data analysis. 
However, Medeiros and Medeiros \cite{medeiros2002} warn that over-reliance on 
simulations may disconnect students from authentic experimental practice, 
underscoring the need for balance between virtual and hands-on learning.

Finally, Villani and Carvalho \cite{villani1994} highlight that without 
guided reflection, students often fail to connect experimental procedures to 
theoretical concepts, which hinders meaningful conceptual change. These 
studies suggest that undergraduate physics education must move beyond 
prescriptive lab manuals and integrate deeper inquiry, explicit treatment of 
uncertainty, and diverse instructional tools to foster robust experimental 
competence.

\section{Experimental setup}

Hydrostatics studies fluids at rest and the forces acting on them. In this 
experiment, we analyze the hydrostatic forces exerted on a submerged object 
to determine the density of water using the principles of buoyancy. The setup 
consists of a graduated cylinder filled with water, a digital scale, and a 
metal bar suspended by an adjustable support. By recording variations in mass 
and volume as the bar is gradually submerged, we can quantify the buoyant 
force exerted by the liquid.

\begin{figure}[h]
    \centering
    \includegraphics[width=1\textwidth]{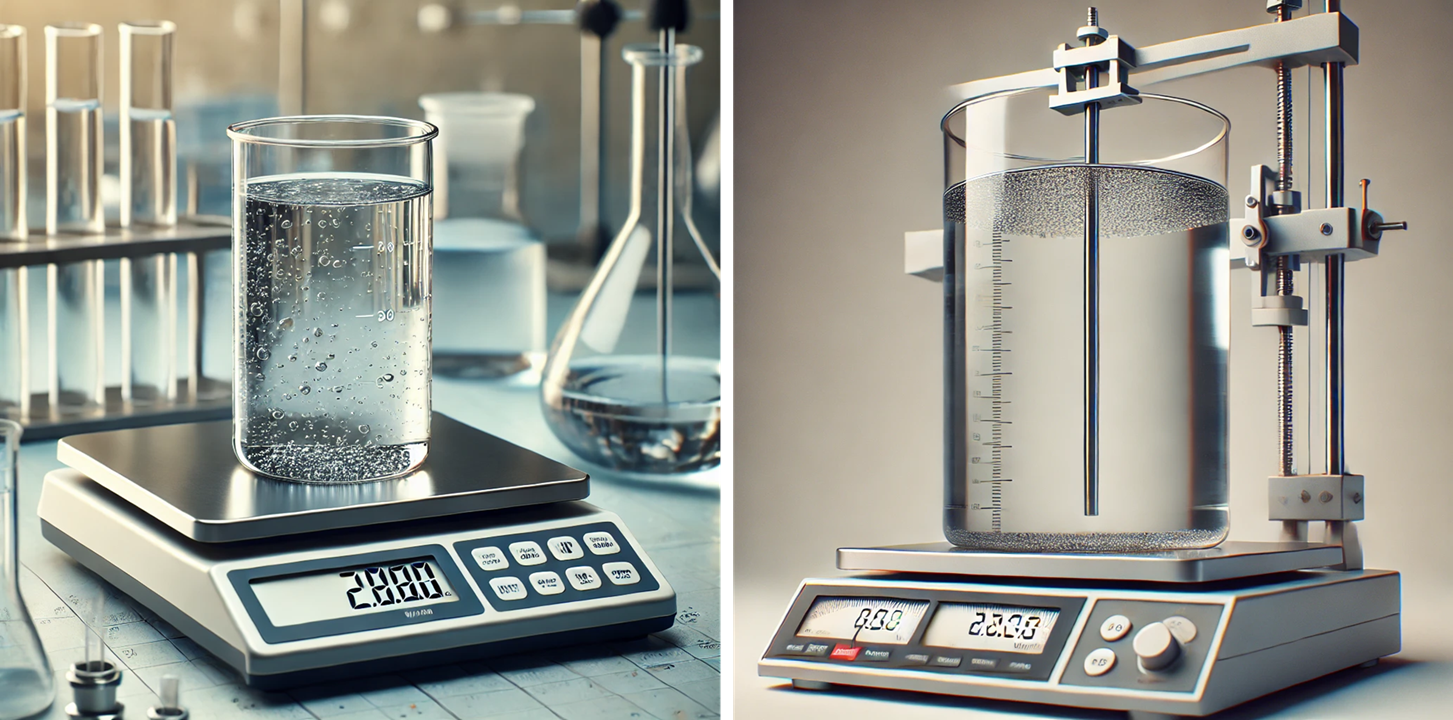}
    \caption[Experimental setup for determining water density.]{Experimental 
setup for determining the density of water using hydrostatic principles. (left
) Initial setup: A container filled with water is placed on a digital scale, 
measuring the total weight of the container and the liquid. right) Modified 
setup: A metal rod is suspended by an apparatus and partially submerged in 
the water. The system demonstrates the buoyant force exerted by the liquid on 
the rod, resulting in changes to the scale's reading. By analyzing these 
variations, the density of the liquid can be experimentally determined using 
Archimedes' principle. ChatGPT-4o generated both images.}
    \label{fig:expsetup}
\end{figure}

The experimental apparatus, depicted in Fig.\,\ref{fig:expsetup}, consists of 
two distinct stages. In the first stage, the metal bar is positioned outside 
the liquid, held in place by a support that ensures it does not interact with 
the fluid. The tension in the support balances the weight of the bar, keeping 
it in equilibrium. In this state, the scale measures the combined mass of the 
graduated cylinder and the liquid, denoted by $M_0$. The initial liquid 
volume is $V_0$, providing a reference measurement.

In the second stage, the bar is partially submerged in the liquid. As the bar 
is lowered using the adjustable support, it displaces a volume of fluid, now 
represented as $V_d$. According to Archimedes' principle, the fluid exerts an 
upward buoyant force $E$ on the submerged portion of the bar. Due to Newton's 
third law, action and reaction, the liquid also experiences an equal and 
opposite force, which alters the scale reading. Consequently, the new mass 
reading on the scale is $M > M_0$ due to the reaction force acting on the 
liquid. This setup allows us to quantify the buoyant force by analyzing the 
variations in mass and volume readings as the bar is submerged.

The materials used in this experiment include a graduated cylinder to measure 
liquid displacement, a scale to record mass variations, metal bars of 
different materials and cross-sections, water as the working fluid, and 
support with a crank for controlled vertical movement of the metal bar.

The following steps are followed: First, the mass of the empty container is 
measured and denoted as $M_R$. The scale's precision is checked, and the most 
minor measurable division is noted. Ensure the support and scale are leveled 
for accurate readings. The liquid's initial volume $V_0$ in the graduated 
cylinder is recorded. The liquid level is adjusted to ensure that the bar can 
be fully submerged without overflowing.

For data collection, the values of $M_0$ and $V_0$ are measured with the 
metal bar completely outside the liquid. The bar is lowered incrementally 
into the liquid using the crank, displacing a volume $V_d$ each time the 
experiment is executed. The new mass, $M_1$, and volume, $V_1$, are recorded. 
This process is repeated for additional measurements $(M_2, M_3, \dots)$ and $
(V_2, V_3, \cdots)$ while ensuring that the bar remains suspended and does 
not touch the graduated cylinder. The students performing the data 
acquisition must record the measured values of Mass M and volume V in a 
proper table, along with their respective measurement errors, $\ sigma_M$ for 
the mass and $\sigma_V$ for the volume.

The experiment is conducted using two different metal bars, the objective of 
using two metal bars is to bring to the attention of the students 
experimenting that the calculation of the water density does not depend on 
the type of material of the two rods, but only on the submerged volume inside 
the liquid in the recipient, since from Archimede's Principle, the buoyant 
force only depends on the liquid density and the displaced volume of liquid. 
For the second bar, measurements are taken only for volumes equal to or 
greater than the final measured volume of the first bar. This setup enables 
direct experimental verification of Archimedes' principle by relating mass 
variations to the displaced volume of liquid.

The collected data must be processed, refined, and analyzed by the students 
to calculate water density using simple linear regression. This involves 
using the angular coefficients of the estimated line to determine the value 
of the water density.

It must be disclaimed that the two images in Fig.\,\ref{fig:expsetup} were 
created using Generative Artificial Intelligence (ChatGPT-4o).

To ensure precise control over the displacement of the metal rod into the 
water, the experimental setup incorporated an adjustable support mechanism 
coupled with a fine-threaded crank system. The crank allowed for smooth, 
incremental lowering of the rod, minimizing sudden movements and vibrations 
that could affect the stability of the measurements. Each crank turn 
corresponded to a calibrated vertical displacement, enabling the operator to 
adjust the rod's immersion depth with high reproducibility. Additionally, the 
student performing the experimental measurements must use the locking 
mechanism on the adjustable support to hold the rod in place during mass and 
volume readings, ensuring no additional movement occurs during data 
acquisition. This system also played a crucial role in maintaining the rod's 
alignment, preventing it from contacting the walls of the container. Such 
contact could introduce unwanted tangential and normal forces due to friction 
with the recipient’s surface, leading to measurement artifacts on the scale. 
By avoiding these forces, the setup helped preserve the accuracy and 
reliability of the mass readings during the experiment.

Two different metal rods were intentionally used to help students recognize a 
key aspect of Archimedes' principle: that calculating the fluid’s density 
does not depend on the geometrical properties or the material composition of 
the submerged object. According to Archimedes' principle, the buoyant force 
acting on a fully or partially submerged object depends solely on the density 
of the fluid and the volume of the displaced liquid, regardless of the 
object's shape, density, or material. Using rods with distinct densities and 
geometries, students can experimentally verify that the calculated value of 
the water’s density remains the same.

Below in Fig.\,\ref{fig:density_experiment} is a schematic illustration of 
all the elements used in the experiment to determine the water density value. 
Elements on the schematic figure are: (A) glass container with known total 
mass $M_0$ (water + container) and volume $V_0$ of water inside; (B) metal 
rod with known mass $M_R$; (C) crank for precise lowering of the metal rod; (D
) scale; (E) scale measurement arm.

A disclaimer must be made that the image in Fig.\,\ref{fig:density_experiment}
 was created using Generative Artificial Intelligence (ChatGPT-4o), and the 
author altered the resulting image to include the labeled elements (A), (B), (
C), (D), (E) with the purpose to describe the experimental setup better. It 
is worth noting that, despite some evident design flaws, the image is 
reasonably decent and can serve as a visual aid for readers.

The student must first annotate the initial volume, $V_0$, of water inside 
the recipient, as indicated by the walls of the container, in milliliters. 
Then, measure the mass $M_0$ of the container and the water (A), excluding 
the immersed metal rod (B). Then the students must proceed to use the crank (C
) to lower the metal rod (B) carefully and slowly so it does not spill any 
water, and to avoid the metal rod touching the walls of the container to 
prevent other forces from appearing due to friction and making the mass 
measurements less precise. After the rod is immersed, the students must 
anotate the new volume $V$ in the markers on the wall of the container, now 
with the added displaced volume $\delta V = V - V_0$ of water due to the 
immersion of the rod, and then use the scale measurement arm (E) to annotate 
the new mass measurement $M$ now containing the mass of the immersed rod in 
the water and noting that Archimedes' principle states that the buoyant 
forces on immersed objects in liquids are proportional to the displaced 
fluid's weight by the submerged object's volume.

\begin{figure}[h]
    \centering
    \includegraphics[width=0.65\textwidth]{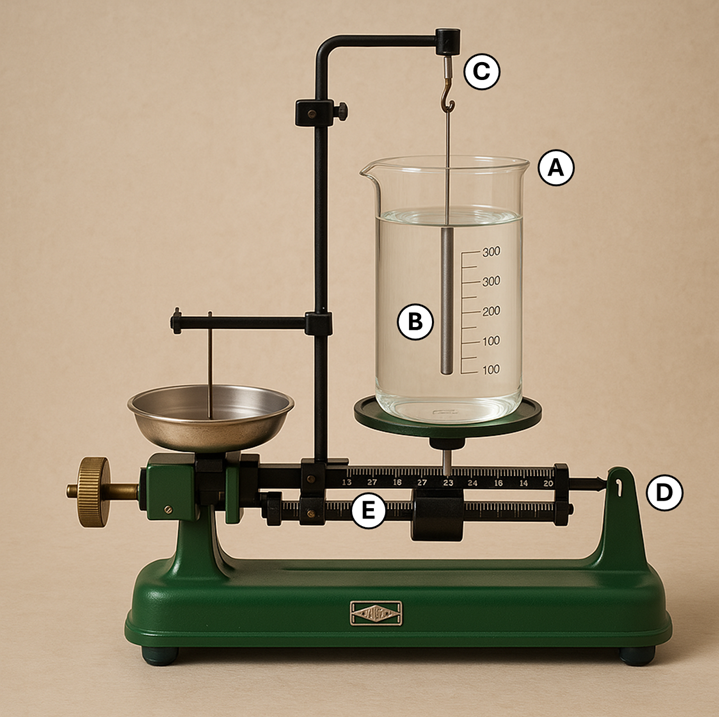}
    \caption[schematic figure of the experiment apparatus]{
        Experiment to determine water density using Archimedes' principle. 
Elements on the schematic figure are: 
        (A) glass container with known total mass $M_0$ (container + water) 
and volume $V_0$ of water; (B) metal rod with known mass $M_R$; (C) crank for 
precise lowering of the rod; (D) scale;
        (E) scale measurement arm.
    }
    \label{fig:density_experiment}
\end{figure}

\section{Pedagogical Approach and Pitfalls}

This section presents several caveats that students should avoid when 
performing the water density determination to ensure accurate data points for 
analysis. The group of students must follow the procedure listed below.

\begin{itemize}
    \item[-] Using the data obtained for water, create a table containing the 
quantities $M$ and $V$, along with their respective uncertainties.
    
    \item[-] Initially, identify in the table which parameters are obtained 
directly and indirectly from the experiment.
    
    \item[-] In the report, show how the results of the indirect measurements 
and their respective uncertainties were determined—for example, the error on 
the Ordinary Least Squares parameters estimators.
    
    \item[-] Use the equations from the theoretical framework to determine 
the equation of the line $M = a V + b$ and then perform a linear fit using 
the experimental data to determine the values of the slope and intercept 
coefficients. 
    
    \item[-] With slope and intercept values or the linear model, indirectly 
determine the water density value and calculate the estimate's error. 

    \item[-] Anotate the values of $a \pm \delta a$ for the slope and $b \pm 
\delta b$ for the intercept in tables in the report.
\end{itemize}

To avoid pitfalls, students must be attentive to certain caveats in the 
experimental procedure.

\begin{itemize}
    \item[-] The same student must perform the same procedure to reduce 
errors since each has a different sight, height, or manner of doing the 
measurements. 

    \item[-] The group of students must be organized and methodical to write 
down the data as soon as the measurement is performed.

    \item[-] Watch out for the significant numbers of each measurement on the 
mass and the volume. In some experiments, the setup may be intentionally `old 
school'. For example, using an old scale instead of a precision scale.

    \item[-] Be careful with the crank when lowering the metal rod. If an 
angle is formed with the vertical, tangential forces may appear, and an 
experimental error may affect the final calculated value for the water density.

    \item[-] Do not let the metal rod touch the sides of the container for 
the same reason as the last item. Tangential forces may arise due to the 
contact between the rod and the recipient wall.

    \item[-] Be aware of the dimensional analysis. The water density is 0.997 
g/mL at 25 degrees Celsius.
\end{itemize}

There are some caveats that students must be aware of regarding the physical 
interpretation of the experimental results.

\begin{itemize}
    \item[-] What model is the best choice to reduce errors if it is either $
M \times V$ or $V \times M$? And why is that so?

    \item[-] How to calculate the linear regression estimator errors that fit 
the data with the best line.

    \item[-] What is the physical interpretation of $M_R = M_0 - \rho V_0$.

    \item[-] Why is the slope calculation in a millimeter paper less accurate 
than ordinary least squares?

    \item[-] How to properly propagate errors and estimate experimental 
mistakes.
\end{itemize}

\section{Theoretical Framework}

In this section i present a theoretical framework, with the derivation of 
the equation that relates the experimental measurements of mass for the 
system metal rod + glass container + water $(M)$ and the total volume $V$ 
from the initial water volume $V_0$ and the displaced water volume by the 
immersion of the metal rod in the liquid.

\begin{figure}[h]
    \centering
    \includegraphics[width=0.5\textwidth]{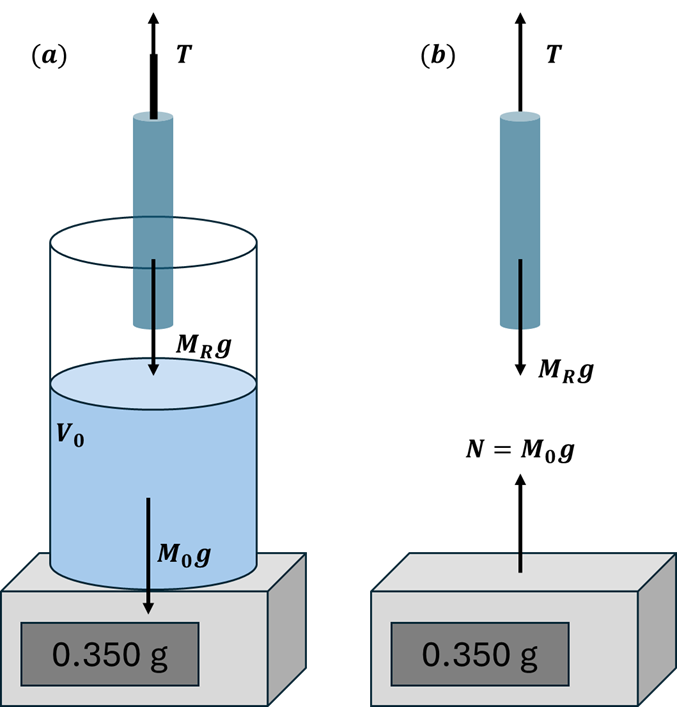}
    \caption[Newton's law and Archimedes' principle.]{Free body diagram of 
forces acting on the experimental setup composed by the glass container, the 
metal rod holder by the crank, the water inside the container, and the scale, 
while the metal rod is not yet immersed in the liquid. Image (a) on the left 
shows the experimental setup, and image (b) on the right shows the free body 
diagram of acting forces on the apparatus.}
    \label{fig:newtonlaw1}
\end{figure}

Fig.\,\ref{fig:newtonlaw1} shows a free body diagram illustration for the 
configuration where the metal rod is not yet immersed in the liquid. The 
illustration shows the acting forces on the system composed of the glass 
container, the water inside the container, the scale, the metal rod, and 
the crank holding the metal rod. Image (a) on the left shows the experimental 
setup with all the experimental elements and the acting forces, and image (b) 
on the right depicts only the free body diagram of acting forces on the 
experimentalsetup.

For the configuration showned in Fig.\,\ref{fig:newtonlaw1}, the metal rod is 
not yet immersed in the liquid, hence the only two acting forces on the metal 
rod are the tension $T$ acting on the crank support that holds the rod, and the 
weight of the metal rod given by $(M_{R} g)$. Therefore, the mass $M_0$ 
measured by the scale is calculated by considering the reaction force $N$ in 
opposition to the weight $M_0 g$ of the system, which includes the glass 
container and the water.

\begin{figure}[h]
    \centering
    \includegraphics[width=0.5\textwidth]{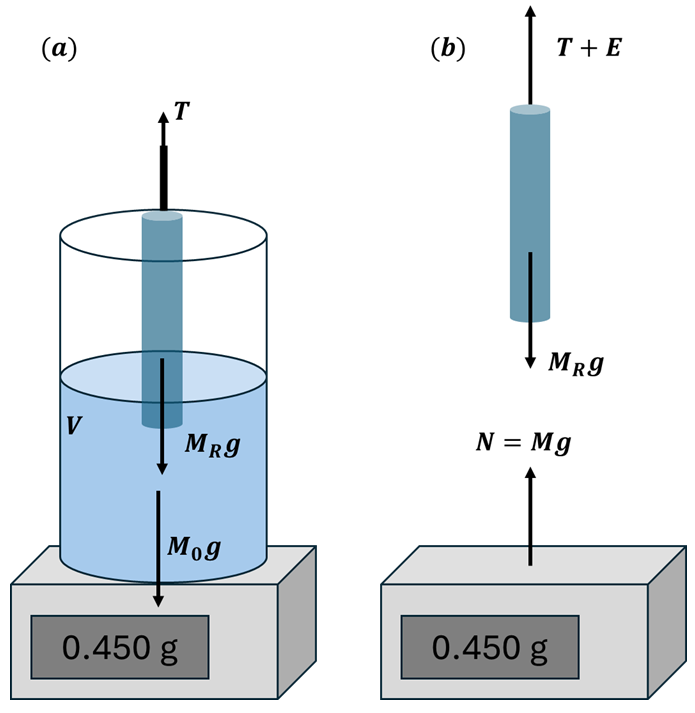}
    \caption[Newton's law and Archimedes' principle.]{Free body diagram of 
forces acting on the experimental setup composed by the glass container, the 
metal rod holder by the crank, the water inside the container, and the scale, 
while the metal rod is immersed in the liquid. Image (a) on the left shows 
the experimental setup, and image (b) on the right shows the free body 
diagram of acting forces on the apparatus.}
    \label{fig:newtonlaw2}
\end{figure}

Fig.\,\ref{fig:newtonlaw2} below is very similar to Fig.\,\ref{fig:newtonlaw1}. 
Still, in a different configuration, the metal rod was now lowered by the 
crank and displaced a volume $ V-V_0$ of water inside the glass container. 
Hence, the new mass $M$ measured by the scale is given by the original mass $M
_0$ plus the displaced volume of water. Image (a) on the left depicts the new 
configuration with the metal rod lowered inside the liquid, and image (b) on 
the right depicts the free body diagram of forces acting on the system. The 
new normal reaction acting on the scale is $N = M g$, where $ M-M_0$ is the 
mass of the displaced volume of water by the partially submersed crank. Now 
the acting forces on the metal rod are the tension $T$ by the crank, the 
weight $M_R g$, and the buoyant force $E = \rho (V-V_0)g$ given by 
Archimedes' principle.

In the scenario where the metal bar is not yet immersed in the water, the 
scale only reads the reaction of the normal force on the container + liquid 
system
\be
F_0 = M_0 g
\ee
where $F_0$ is the force acting on the scale, $M_0$ is the container's mass 
plus the liquid's mass, and $g$ is the acceleration due to gravity. When a 
metallic bar is partially immersed in the liquid, forces begin to act on both 
the liquid and the bar. In the static situation, only pressure forces 
contribute to the resultant force since the force due to the viscosity of the 
liquid depends on the relative velocity between the bar and the fluid. The 
sum of the pressure forces that a liquid exerts on a solid is called the 
buoyant force, and Archimedes' Principle gives it 
\cite{halliday2010physics, nussenzveig_fisica_basica_vol2}:
\be
E = \rho V_d g \label{archprinc}
\ee
where $\rho$ is the density of the liquid, and $V_d$ is the volume of liquid 
displaced by the solid. In this situation, the buoyant force acts upwards, 
counteracting the force that pushes the metal bar out of the liquid. The 
reading on the scale is now $M > M_0$ since a Buoyant force is acting on the 
system. The resultant force is now $F = Mg$. Newton's second law applied to 
the liquid + container system results in the following expression
\be
Mg = E + M_0 g
\ee
which can be read as the following expression
\be
E = (M - M_0) g \label{buo1}
\ee

Inserting Eq.\,\eqref{archprinc} in Eq.\,\eqref{buo1}, and noticing that the 
dislocated volume on the container is given by $V_d = V-V_0$, where $V$ is 
the metal bar volume immersed in the liquid, results 
\be
E = (V - V_0) \rho g \label{buo2}
\ee
Eq.\,\eqref{buo1} and Eq.\,\eqref{buo2} both represent the buoyancy force, so 
they must be equal
\be
(V-V_0) \rho g = (M - M_0) g \label{buo3}
\ee
And we can put Eq.\,\eqref{buo3} in the following manner
\be
M = \rho V + (M_0 - \rho V_0) \label{modelm}
\ee
or also in the following manner
\be
V = \frac{M}{\rho} + \left(V_0 - \frac{M_0}{\rho} \right) \label{modelv}
\ee
The theoretical model predicts that the buoyant force does not depend on any 
property of the solid, but only on the volume of the object immersed in the 
liquid, as seen in Eq.\,\eqref{archprinc}. Eq.\,\eqref{modelm} and Eq.\,\eqref
{modelv} both represent the same model and show a linear relationship between 
$M$ and $V$, of the form $y = ax + b$, where $a$ is the slope and $b$ is the 
linear coefficient. However, it is worth noting that statistically, there are 
differences between these two models.

\section{Statistical tools and error analysis}

Linear regression is a fundamental tool in experimental physics, enabling 
researchers and students to derive physical constants and model relationships 
between measured variables. Bevington and Robinson \cite{bevington2003} offer 
one of the most comprehensive treatments of linear regression within the 
context of experimental data analysis, emphasizing the importance of least-
squares fitting in interpreting measurements. Taylor \cite{taylor1997} 
complements this approach by focusing on the role of uncertainties, guiding 
students on integrating error analysis into regression results to assess the 
reliability of their conclusions. Hill \cite{hill2017} provides a practical 
laboratory manual that introduces linear regression in introductory physics 
labs, helping students understand the computational and conceptual aspects of 
data fitting. Meanwhile, Cleveland \cite{cleveland1993} addresses regression 
from a data visualization perspective, highlighting how graphical 
representations can aid in interpreting experimental trends. Holmes and 
Wieman \cite{holmes2018} critique the superficial use of regression in many 
labs, warning that students often apply linear fits without fully engaging 
with their scientific meaning or understanding the propagation of uncertainty.

Linear regression is widely used to model the relationship between two 
variables when expected to follow a linear trend. Consider a set of data 
points $(x_i, y_i)$ for $i=1,2,...,N$. The model assumes the relationship:
\be
y = ax + b\,,
\ee
where $m$ is the slope and $b$ is the intercept. The slope and intercept can 
be derived from first principles by minimizing the sum of squared residuals
\be
S = \sum_{i=1}^{N} (y_i - mx_i - b)^2\,.
\ee
Taking the partial derivatives of $S$ concerning $m$ and $b$ and setting them 
to zero yields the standard equations. Solving them yields
\be
a = \frac{N \sum x_i y_i - \sum x_i \sum y_i}{N \sum x_i^2 - (\sum x_i)^2},
\ee
\be
b = \frac{\sum y_i - a \sum x_i}{N}.
\ee
These formulas are commonly used in experimental physics to fit data to a 
linear model. However, in many physical experiments, the relationship between 
variables is nonlinear, such as
\be
y = a \exp(bx).
\ee
This can be linearized by taking the natural logarithm:
\be
\ln(y) = \ln(a) + bx,
\ee
allowing the use of linear regression on $\ln(y)$ versus $x$ to estimate $b$ 
and $\ln(a)$. Another example is a power-law relationship:
\be
y = ax^n,
\ee
which can be linearized as
\be
\ln(y) = \ln(a) + n \ln(x).
\ee
Accurate analysis in experimental physics requires understanding how 
uncertainties propagate through calculations. For a function $f$ depending on 
variables $x$ and $y$
\be
f = f(x, y),
\ee
the uncertainty in $f$, denoted $\sigma_f$, is given by
\be
\sigma_f = \sqrt{\left( \frac{\partial f}{\partial x} \sigma_x \right)^2 + 
\left( \frac{\partial f}{\partial y} \sigma_y \right)^2},
\ee
where $\sigma_x$ and $\sigma_y$ are the uncertainties in $x$ and $y$, 
respectively. For example, for $f = xy$
\be
\sigma_f = f \sqrt{\left( \frac{\sigma_x}{x} \right)^2 + \left( \frac{\sigma_y
}{y} \right)^2}.
\ee
This propagation formula is crucial for evaluating the final uncertainty in 
calculated physical quantities. For further discussion, readers can consult 
Bevington and Robinson \cite{bevington2003}, Taylor \cite{taylor1997}, and 
Hill \cite{hill2017}, who provide foundational insights into both linear 
regression and uncertainty analysis.

In summary, linear regression and error propagation are key to drawing 
reliable conclusions from experimental data. Mastering these techniques 
allows physicists to interpret trends, validate models, and quantify the 
confidence in their results.

\section{Data Analysis and Discussion}

This section presents a thorough discussion of the analysis of the 
experimental data. Present how the data should be organized in a table with 
the values for the pairs $(M_i,V_i)$ with their respective errors $\pm \delta 
M_i$ and $\pm\delta V_i$. A discussion is presented on how a model $M \times V
$ is more appropriate than a model $V \times M$ due to error propagation. A 
model $M \times V$ requires a smaller margin of error for statistical 
confidence.

Consider the experimental setup consisting of a graduated container partially 
filled with a liquid of density $\rho_{\text{liquid}}$ and a metallic bar 
that can be gradually immersed in the liquid, as shown in 
Fig.\,\ref{fig:expsetup} and Fig.\,\ref{fig:density_experiment}, then The key 
variables are defined as follows:

\begin{itemize}
    \item[-] $M_0$: Mass reading on the scale when the bar is outside the 
liquid. It is the mass of the system, which includes both the liquid and the 
container. It can be directly measured using the scale.
    \item[-] $M$: Mass reading when the bar is partially immersed. It can be 
measured experimentally. It is formed by the system liquid, a container, and 
a partially submerged metal rod.
    \item[-] $V_0$: Initial volume of the liquid before submersion of the 
metal rod. It can be directly measured.
    \item[-] $V$: Volume of the liquid after submersion. It is the volume $V_0
$ added by the dislocated volume $V_d$. It can be directly measured 
experimentally.
    \item[-] $V_{d}$: Volume of the liquid displaced by the submerged part of 
the bar. It can only be calculated with the formula $V_d = V - V_0$.
    \item[-] $g$: Acceleration due to gravity. It cannot be directly measured 
but does not play a key role in this experiment; it only appears due to 
Newton's laws.
    \item[-] $E$: Buoyant force acting on the submerged portion of the bar.
    \item[-] $M_R$: It is the intercept of the model $M = a V + M_R$, defined 
by $M_R = M_0 - \rho V_0$. It cannot be measured directly; it is only 
calculated.
\end{itemize}

The table below presents the measured mass and volume values of a submerged 
object to study buoyancy and fluid properties in an experimental setup. Each 
measurement includes its associated uncertainty, denoted as $\delta M_i$ for 
mass and $\delta V_i$ for volume, which accounts for instrumental precision 
and experimental variations. Additionally, the relative uncertainties, $\frac{
\delta M_i}{M_i}$ and $\frac{\delta V_i}{V_i}$, are provided to quantify the 
accuracy of the measurements. 

\begin{table}[ht]
    \centering
    \renewcommand{\arraystretch}{1.20} 
    \begin{tabular}{c p{2.5cm} p{1cm} p{3cm} p{1cm}}
        \hline
        \rule{0pt}{15pt} 
        $i$ & $(M_i \pm \delta M_i) \ g$ & $\frac{\delta M_i}{M_i}$ & $(V_i 
\pm \delta V_i) \ ml$ & $\frac{\delta V_i}{V_i}$ \\ \vspace{-12pt} \\
        \hline 
        1  & 208.12 $\pm$ 5.20  & 0.025 & 110.00 $\pm$ 5.50  & 0.050 \\
        2  & 217.37 $\pm$ 5.45  & 0.025 & 120.00 $\pm$ 6.00  & 0.050 \\
        3  & 228.34 $\pm$ 5.71  & 0.025 & 130.00 $\pm$ 6.50  & 0.050 \\
        4  & 241.61 $\pm$ 6.05  & 0.025 & 140.00 $\pm$ 7.00  & 0.050 \\
        5  & 251.05 $\pm$ 6.28  & 0.025 & 150.00 $\pm$ 7.50  & 0.050 \\
        6  & 262.01 $\pm$ 6.55  & 0.025 & 160.00 $\pm$ 8.00  & 0.050 \\
        7  & 272.14 $\pm$ 6.80  & 0.025 & 170.00 $\pm$ 8.50  & 0.050 \\
        8  & 278.10 $\pm$ 6.95  & 0.025 & 180.00 $\pm$ 9.00  & 0.050 \\
        9  & 290.44 $\pm$ 7.26  & 0.025 & 190.00 $\pm$ 9.50  & 0.050 \\
        10 & 297.54 $\pm$ 7.44  & 0.025 & 200.00 $\pm$ 10.00 & 0.050 \\
        \hline
    \end{tabular}
    \caption{Experimental Data Table}
    \label{tab:experimental_data}
\end{table}

From Table \ref{tab:experimental_data}, it is evident that the relative 
uncertainties in both mass and volume measurements remain consistent, with $
\frac{\delta M_i}{M_i} = 0.025$ and $\frac{\delta V_i}{V_i} = 0.050$ across 
all data points. This consistency ensures that the error propagation in the 
regression analysis is well-controlled. The regression computation 
incorporated the experimental uncertainties to provide a more robust 
estimation of the parameters.

To estimate the density of water from the experimental data, students can 
determine the slope of the mass-volume relationship using a simple graphical 
method on millimeter paper. A straight-line approximation can be drawn 
through the data points by plotting the mass $M$ against the volume $V$. The 
slope of this line, which corresponds to the density, can be estimated using 
the fundamental definition from differential calculus:
\be
a = \frac{\Delta M}{\Delta V} = \frac{M_2 - M_1}{V_2 - V_1}
\ee
where $(V_1, M_1)$ and $(V_2, M_2)$ are two points chosen from the 
experimental data. For instance, selecting the points $(V_1 = 110.0, M_1 = 210
.81)$ and $(V_2 = 200.0, M_2 = 300.78)$ from the experimental table, we 
compute the slope as:
\be
a = \frac{300.78 - 210.81}{200.0 - 110.0} = \frac{89.97}{90.0} = 0.9997 \, 
\text{g/mL}.
\ee
While this method estimates the density, it is susceptible to the specific 
points chosen for analysis. Ideally, the best-fit line for the data should be 
obtained through an Ordinary Least Squares regression, which minimizes the 
sum of squared residuals, given by:
\be
e_i = M_i - \hat{a} V_i - \hat{b},
\label{residue}
\ee
Where $\hat{a}$ and $\hat{b}$ are the slope and intercept of the best-fit 
line, the best estimator parameters for the best line selected by the 
Ordinary Least Squares method. This approach minimizes the overall error 
across all data points, rather than relying solely on two chosen points. In 
contrast, manually selecting points introduces significant variability, as 
minor fluctuations in measurement values can lead to disproportionately large 
errors in the estimated slope.

By employing Ordinary Least Squares regression, students can more accurately 
determine the density of water while accounting for the inherent 
uncertainties in experimental data. Though useful for a rough approximation, 
the graphical method is prone to errors that statistical regression 
techniques can systematically reduce.

When selecting a model to determine the density of water, one must consider 
the mathematical implications of choosing either the $M \times V$ model, 
where mass is expressed as a function of volume, or the $V \times M$ model, 
where volume is described as a function of mass. The choice significantly 
affects the accuracy of density estimation due to differences in how errors 
propagate. For the $M \times V$ model, the relationship is given by:
\be
M = \rho V + (M_0 - \rho V_0),
\ee
where the slope of the regression line directly corresponds to the water 
density, $\rho$. The uncertainty in the estimated slope, $\sigma_a$, is given 
by:
\be
\sigma_a = \frac{\sigma}{\sqrt{\sum_{i=1}^{N} (V_i - \bar{V})^2 } },
\label{erroramxv}
\ee
Where $\sigma$ is the standard deviation of the residuals $e_i$ from Eq.
\,\eqref{residue} can be approximated by a normal distribution and estimated 
from the data in Table \ref{tab:experimental_data}. This model enables a 
straightforward calculation of $\rho$, as the slope of the linear fit 
directly provides it. On the other hand, the $V \times M$ model follows the 
equation:
\be
V = \frac{1}{\rho}M + \left(V_0 -\frac{M_0}{\rho}\right).
\ee
Now, the uncertainty in the estimated slope is given by
\be
\sigma_a = \frac{\sigma}{\sqrt{\sum_{i=1}^{N} (M_i - \bar{M})^2 } },
\label{erroravxm}
\ee
Here, the slope of the regression line is $a = \frac{1}{\rho}$, meaning that 
the density must be obtained by inverting the slope: 
\be
\rho = \frac{1}{a}.
\ee
However, this inversion introduces a more complex error propagation, 
decreasing the uncertainty in $\rho$. The standard error in the density 
estimate becomes:
\be
\sigma_\rho = \left| \frac{d\rho}{da} \right| \sigma_a = \frac{\sigma_a}{a^2}.
\ee
Since the error is magnified by the inverse square of the slope, the $V \times
 M$ model leads to a less significant error in the estimate of $\rho$ 
compared to the $M \times V$ model. Additionally, since the measurement error 
of $M$ is more accurate than the errors in volume, the error estimation of 
the slope $a$ is reduced when $M$ is used in the abcissa instead of $V$. This 
can be seen from Eq.\,\eqref{erroramxv} and Eq.\,\eqref{erroravxm}, where the 
estimate for the slope error $\sigma_a$ is inversely proportional to the sum 
of mean square error for the abscissa values. Using experimental data that 
minimizes the mean square error leads to a better mathematical model, which 
is the case for the mass measurements $M \pm \delta M$. This decreased 
uncertainty makes precise density determination more desirable. Thus, from a 
statistical standpoint, the best approach is to use the $V \times M$ model.

\begin{figure}[ht]
    \centering
    \includegraphics[width=0.9\textwidth]{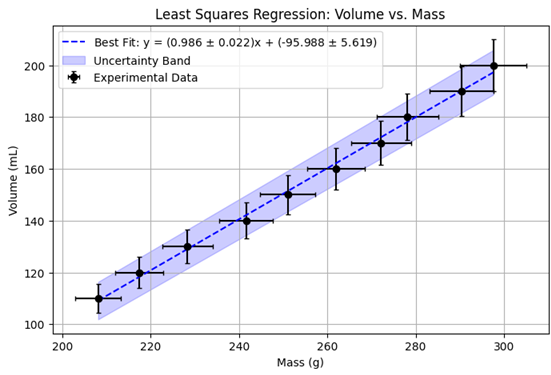}
    \caption{Least Squares Regression: Volume vs. Mass. The plot shows 
experimental data (black markers with error bars) and a linear regression best
-fit line (dashed blue). The equation of the best-fit line is given as \(y = (
0.986 \pm 0.022)x + (-95.988 \pm 5.619) \), where the uncertainties in the 
slope and intercept are provided. The shaded blue region represents the 
uncertainty band around the regression line, indicating the confidence 
interval.}
    \label{fig:volume_vs_mass}
\end{figure}

Fig.\,\ref{fig:volume_vs_mass} displays the data points and the best-fit line 
for the model $V \times M$, where the abscissa values represent the mass 
measurements. The figure displays the error bars for the x-axis and y-axis 
measurements, along with an uncertainty band. 

The estimated value for the water density found for this model was
\be
\rho_{V\times M} \pm \sigma_\rho = 0.986 \pm 0.022
\label{vxm}
\ee
Considering the reference value $(\rho_{\text{ref}})$ for the water density 
at $25^\circ$ as
\be
\rho_{\text{ref}} \pm \sigma_\rho = 0.997 \pm 0.001 \,,
\label{waterref}
\ee
one can calculate the relative discrepancy $(D)$ given by the formula below
\be
D = \left|\frac{x_{\text{ref}} - \bar{x}}{x_{\text{ref}}} \right| \,,
\label{reldisc}
\ee
where $x_{\text{ref}}$ is the reference value of the physical quantity we are 
calculating, and $\bar{x}$ is the calculated value using the mathematical 
model for the experiment. In this case, $x$ is the water density.

From Eq.\,\eqref{waterref}, the measurement interval for the water density 
ranges from $0.996$ to $0.998$. So, the calculated value for the water 
density in Eq.\,\eqref{vxm} is out of the accepted measured interval, and the 
relative discrepancy is
\be
D = \left|\frac{0.997 - 0.986}{0.997} \right| = 1.15 \%\,.
\ee
Since the calculated value for water density falls outside the accepted 
measurement interval, the precision of this calculation needs to be improved. 
It would be necessary to repeat the experiment, paying close attention to the 
measured values to minimize experimental errors.

\begin{figure}[h]
    \centering
    \includegraphics[width=0.9\textwidth]{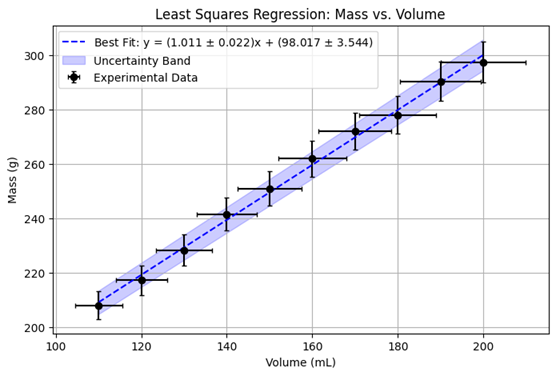}
    \caption{Least Squares Regression: Mass vs. Volume. The plot shows 
experimental data (black markers with error bars) and a linear regression best
-fit line (dashed blue). The equation of the best-fit line is given as \(y = (
1.011 \pm 0.022)x + (98.017 \pm 3.544) \), where the uncertainties in the 
slope and intercept are provided. The shaded blue region represents the 
uncertainty band around the regression line, indicating the confidence 
interval.}
    \label{fig:mass_vs_volume}
\end{figure}

Fig.\,\ref{fig:mass_vs_volume} shows the data points for the model $M \times V
$, the best fitted line using Ordinary Least Squares, and the error bars for 
the measurements of volume (abscissa) and the mass (ordinate), and an 
uncertainty band. 

The estimated value for the water density found for this model was
\be
\rho_{M\times V} \pm \sigma_\rho = 1.011 \pm 0.022\,.
\ee
The model $M \times V$ has a worse precision for the water density 
calculation than the model $V \times M$, considering the same data points. 
This happens due to the larger error in the volume measurements now used as 
an explanatory variable. The value found for this model is also outside the 
accepted measurement interval for the water density value of $0.997 \pm 0.001$
. The relative discrepancy is
\be
D = \left|\frac{0.997 - 1.011}{0.997} \right| = 1.37 \%\,.
\ee

Table \ref{tab:mass_volume_data} contains the experimental data points for 
mass $(M)$ and volume $(V)$, the respective squared errors $(M_i - \bar{M})^2$
 for mass and $(V_i - \bar{V})^2$ for the volume, and experimental errors $
\delta M_i$ and $\delta V_i$ for each of the measurements.

To understand how the precision of the measurements affects the mathematical 
models results and the calculated value for the water density, divide the 
estimated error $\sigma_a^{(M\times V)}$ given by Eq.\,\eqref{erroramxv} for 
the model $M \times V$ by the error $\sigma_a^{(V\times M)}$ from model $V 
\times M$ in Eq.\,\eqref{erroravxm}, considering the same value for the 
standard deviation $\sigma$, then
\be
\frac{\sigma_a^{(M\times V)}}{\sigma_a^{(V\times M)}} = \sqrt{\frac{\sum_{i=1}
^{N} (M_i - \bar{M})^2}{\sum_{i=1}^{N} (V_i - \bar{V})^2}} = 1.01
\ee

\begin{table}[ht]
\centering
\renewcommand{\arraystretch}{1.20}
\begin{tabular}{c c c c c c c}
\hline \rule{0pt}{15pt}
$i$ & $M_i$ [g] & $(M_i - \bar{M})^2 \ [\text{g}^2]$ & $\delta M_i$ [g] & $V_i
$ [ml] & $(V_i - \bar{V})^2 \ [\text{ml}^2]$ & $\delta V_i$ [ml] \\ \vspace{-
14pt} \\
\hline
1 & 208.12    & 2167   & 5.20  & 2025  & 2025   & 5.5 \\
2 & 217.37    & 1391   & 5.45  & 1225  & 1225   & 6.0 \\
3 & 228.34    & 693    & 5.71  & 625   & 625    & 6.5 \\
4 & 241.61    & 171    & 6.05  & 225   & 225    & 7.0 \\
5 & 251.05    & 13     & 6.28  & 25    & 25     & 7.5 \\
6 & 262.01    & 54     & 6.55  & 25    & 25     & 8.0 \\
7 & 272.14    & 305    & 6.80  & 225   & 225    & 8.5 \\
8 & 278.10    & 549    & 6.95  & 625   & 625    & 9.0 \\
9 & 290.44    & 1279   & 7.26  & 1225  & 1225   & 9.5 \\
10 & 297.54   & 1838   & 7.44  & 2025  & 2025   & 10.0 \\
\hline
\end{tabular}
\caption{This table contains the experimental data points $(M,V)$, squared 
errors for mass and volume, and respective experimental errors.}
\label{tab:mass_volume_data}
\end{table}

\section{Pedagogical Discussion}

This experiment is a fundamental exercise in experimental physics, teaching 
students the essential skills required to derive, measure, and analyze 
physical quantities that cannot be directly observed with the available 
apparatus. Determining water density exemplifies how direct measurements of 
mass and volume can be used to estimate an unknown parameter through 
mathematical modeling and data analysis. This process is crucial for students 
to develop a deeper understanding of physical laws and how to translate 
observed phenomena into quantitative models.

A key learning outcome of this experiment is the necessity of deriving 
mathematical equations that describe physical reality. Students must 
establish a theoretical framework that links mass and volume to density, 
collect corresponding data points, and use regression techniques to estimate 
the model parameters. This structured approach fosters a deeper understanding 
of how scientific models are developed and refined, emphasizing that physical 
observables are often not directly measurable but must be inferred from 
empirical data.

Beyond theoretical modeling, students are introduced to statistical methods 
for treating experimental data. Implementing Ordinary Least Squares 
regression is a crucial step in the learning process, as it enables parameter 
estimation by minimizing residual errors. Many students struggle to grasp the 
significance of this method despite its historical relevance dating back to 
Gauss and its continued application in modern data analysis. By working 
through this experiment, students gain firsthand experience applying 
regression to actual data, appreciating its importance in ensuring accurate 
and reliable estimations.

Moreover, this experiment highlights the necessity of computational methods 
in modern physics and engineering. Real-world datasets often contain missing 
values, errors, or inconsistencies, making manual data handling impractical. 
Encouraging students to use programming tools such as Python for data 
analysis fosters a computational mindset, equipping them with indispensable 
skills in today's data-driven scientific landscape. With advancements in 
machine learning and neural networks, data estimation and gap-filling 
techniques have become more sophisticated, and students must be aware of 
these evolving methodologies.

Another critical pedagogical aspect of this experiment is the emphasis on 
graphical representation. In an era where data literacy is increasingly 
essential, students must learn to interpret and construct meaningful 
visualizations. Many struggle with reading tables or understanding simple 
linear relationships between independent and dependent variables. By using 
graphing techniques, students develop a more precise intuition for how one 
physical quantity influences another within a mathematical model. These 
skills are vital in physics and a broad range of STEM disciplines, where data 
visualization plays a crucial role in decision-making and communication.

Finally, an often-overlooked yet fundamental skill in experimental physics is 
the ability to write a structured scientific report. Communicating findings, 
formally and technically, is essential for students pursuing careers in STEM 
fields. The ability to articulate experimental objectives, describe 
methodologies, analyze results, and present conclusions coherently and 
professionally is just as important as experimenting. By emphasizing the 
scientific method in their writing, students refine their ability to document 
and effectively convey their findings, preparing them for future research and 
technical work.

In conclusion, this experiment provides a comprehensive learning experience 
that integrates theoretical modeling, statistical data treatment, 
computational tools, graphical literacy, and scientific communication. By 
engaging with these elements, students develop a well-rounded skill set that 
prepares them for more complex challenges in physics, engineering, and data 
science. Encouraging a rigorous approach to experimental analysis enhances 
their understanding of physical principles and cultivates critical thinking 
and problem-solving abilities essential for any scientific career.

\section{Python class}

The \texttt{WaterDensity} class was implemented in Python to simulate 
experimental data relating mass and volume and to model their relationship 
through linear regression. This Python class was designed to generate 
synthetic datasets and provide analysis tools, including visualization and 
formatted data output for scientific reporting.

The class is initialized with the reference parameters $M_0$, $V_0$, and $\rho
$, respectively, representing the reference mass, reference volume, and fluid 
density. Once initialized, the \texttt{gen\_fake\_data} method can be used to 
simulate experimental data based on a linear model of the form:
\be
M = \rho V + (M_0 - \rho V_0) + \epsilon
\ee
where $\epsilon$ represents random experimental noise, the method outputs a 
dataset including mass ($M$), volume ($V$), and their associated 
uncertainties ($\delta M$ and $\delta V$).

The synthetic data is then analyzed using the \texttt{calculate\_fit} method, 
which applies a least squares linear regression to obtain estimates for the 
slope $a$ and intercept $b$ of the fitted model $M = aV + b$. This method 
also computes the uncertainties in both parameters ($\sigma_a$ and $\sigma_b$)
 and the residual standard error of the fit, $\sigma_y$.

The \texttt{plot\_regression} method generates a plot that displays the 
simulated data points with their corresponding error bars, allowing for a 
clear visualization of the results. The plot also shows the best-fit 
regression line and an uncertainty band derived from the propagated errors in 
the fit parameters.

The data table can be formatted into LaTeX-ready code using the \texttt{format
\_table} method, which produces a structured table displaying $M_i \pm \delta 
M_i$ and $V_i \pm \delta V_i$ values, along with their fractional 
uncertainties.

Finally, the \texttt{export\_latex\_table} method outputs the formatted table 
as LaTeX code. This code can be printed directly to the screen or exported to 
a \texttt{.tex} file for easy integration into LaTeX documents.

The implementation enables the automation of data analysis and reporting for 
experiments that characterize mass-volume relationships, such as determining 
liquid density.

The class was designed for interactive use in Python environments, such as 
Jupyter Notebook. After importing the class with \texttt{from mass\_volume\_
regression import WaterDensity}, the user creates an instance of the class 
and initializes it with values for $M_0$, $V_0$, and $\rho$. The user then 
calls \texttt{gen\_fake\_data} to simulate the dataset. The regression 
analysis is performed using \texttt{calculate\_fit}, and the regression 
results can be visualized with \texttt{plot\_regression}.

Once the data is analyzed, the user can call \texttt{format\_table} to 
prepare the dataset for inclusion in scientific reports. The LaTeX table can 
be printed to the screen or saved as a file using the \texttt{export\_latex\_
table} command.

\begin{lstlisting}[language=Python]
import numpy as np
import pandas as pd
import matplotlib.pyplot as plt

class WaterDensity:
    """
    A class to simulate experimental data for mass vs. volume measurements,
    fit a linear model using least squares regression, and visualize the 
results.
    """

    def __init__(self, M0=300, V0=100, rho=1):
        """
        Initialize the WaterDensity object with reference values.

        Parameters:
        M0 : float - Reference mass (g)
        V0 : float - Reference volume (mL)
        rho : float - Density (g/mL)
        """
        self.M0 = M0
        self.V0 = V0
        self.rho = rho
        self.df = None  # DataFrame to store generated data
        self.fit_results = None  # Dictionary to store regression results

    def gen_fake_data(self, n_points=10, V_min=110, V_max=200, err_M=0.025, 
err_V=0.10, noise=5):
        """
        Generate synthetic mass vs. volume data with uncertainties.

        Parameters:
        n_points : int - Number of data points
        V_min : float - Minimum volume value
        V_max : float - Maximum volume value
        err_M : float - Relative uncertainty in mass
        err_V : float - Relative uncertainty in volume
        noise : float - Random noise added to mass values

        Returns:
        pd.DataFrame containing mass, volume, and their uncertainties.
        """
        # Generate volume values evenly spaced between V_min and V_max
        V_values = np.linspace(V_min, V_max, n_points)
        # Compute mass values using a linear relationship + noise
        M_values = self.rho * V_values + (self.M0 - self.rho * self.V0) + 
np.random.uniform(-noise, noise, n_points)
        # Create a dataframe with uncertainties in mass and volume
        self.df = pd.DataFrame({
            'M': M_values,
            'sigma_M': M_values * err_M,
            'V': V_values,
            'sigma_V': V_values * err_V     
        })
        return self.df

    def calculate_fit(self):
        """
        Perform least squares regression (Mass vs. Volume) on generated data.

        Returns:
        pd.DataFrame and dict containing slope, intercept, uncertainties, and 
predictions.
        """
        if self.df is None:
            raise ValueError("No data available. Please run gen_fake_data() 
first.")
        
        # Extract mass and volume data
        V_values = self.df['V'].values
        M_values = self.df['M'].values
        N = len(V_values)

        # Compute necessary sums for least squares formulas
        sum_x = np.sum(V_values)
        sum_y = np.sum(M_values)
        sum_x2 = np.sum(V_values ** 2)
        sum_xy = np.sum(V_values * M_values)

        # Calculate slope (a) and intercept (b)
        D = N * sum_x2 - sum_x ** 2
        a = (N * sum_xy - sum_x * sum_y) / D
        b = (sum_y - a * sum_x) / N

        # Predicted mass values from regression
        y_pred = a * V_values + b

        # Calculate residual standard error
        sigma_y = np.sqrt(np.sum((M_values - y_pred) ** 2) / (N - 2))

        # Calculate uncertainty in slope and intercept
        Sxx = sum_x2 - (sum_x ** 2) / N
        sigma_a = sigma_y / np.sqrt(Sxx)
        x_bar = sum_x / N
        sigma_b = sigma_y * np.sqrt(1 / N + (x_bar ** 2) / Sxx)

        # Store results
        self.fit_results = {
            'M0': self.M0,
            'V0': self.V0,
            'rho': self.rho,
            'a': a,
            'sigma_a': sigma_a,
            'b': b,
            'sigma_b': sigma_b,
            'y_pred': y_pred,
            'sigma_y': sigma_y,
        }
        return pd.DataFrame([self.fit_results]), self.fit_results

    def plot_regression(self):
        """
        Plot experimental data with error bars, regression line, and 
uncertainty band.
        """
        if self.df is None or self.fit_results is None:
            raise ValueError("You must generate data and calculate fit first.")
        
        # Extract variables and errors
        V_values = self.df['V'].values
        M_values = self.df['M'].values
        V_err = self.df['sigma_V'].values
        M_err = self.df['sigma_M'].values

        # Regression results
        y_pred = self.fit_results['y_pred']
        a = self.fit_results['a']
        b = self.fit_results['b']
        sigma_a = self.fit_results['sigma_a']
        sigma_b = self.fit_results['sigma_b']

        # Sort for plotting a smooth regression line
        sort_idx = np.argsort(V_values)
        V_sorted = V_values[sort_idx]
        y_pred_sorted = y_pred[sort_idx]

        # Propagate uncertainties to compute the uncertainty band
        uncertainty = np.sqrt((V_sorted * sigma_a) ** 2 + sigma_b ** 2)
        y_upper = y_pred_sorted + uncertainty
        y_lower = y_pred_sorted - uncertainty

        # Plot experimental data with error bars
        plt.figure(figsize=(8, 5))
        plt.errorbar(V_values, M_values, xerr=V_err, yerr=M_err,
                     fmt='o', color='black', capsize=2, label="Experimental 
Data")
        # Plot regression line
        plt.plot(V_sorted, y_pred_sorted, linestyle='--', color='b',
                 label=f"Best Fit: y = ({a:.4f} \\pm {sigma_a:.4f})x + ({b:.2f
} \\pm {sigma_b:.2f})")
        # Plot uncertainty band
        plt.fill_between(V_sorted, y_lower, y_upper, color='blue', alpha=0.2,
                         label='Uncertainty Band')
        plt.xlabel("Volume (mL)")
        plt.ylabel("Mass (g)")
        plt.title("Least Squares Regression: Mass vs. Volume")
        plt.grid(True)
        plt.legend()
        plt.show()

    def format_table(self):
        """
        Format the dataframe into a LaTeX-style table with uncertainties.

        Returns:
        pd.DataFrame formatted with \\pm symbols and fractional uncertainties.
        """
        if self.df is None:
            raise ValueError("No data to format. Please run gen_fake_data() 
first.")
        
        return pd.DataFrame({
            "$(M_i \\pm \\delta M_i) \\ g$": [f"{M:.3f} \\pm {sigma_M:.3f}" 
for M, sigma_M in zip(self.df["M"], self.df["sigma_M"])],
            "$\\frac{\\delta M_i}{M_i}$": [f"{(sigma_M / M):.3f}" for M, sigma
_M in zip(self.df["M"], self.df["sigma_M"])],
            "$(V_i \\pm \\delta V_i) \\ ml$": [f"{V:.3f} \\pm {sigma_V:.3f}" 
for V, sigma_V in zip(self.df["V"], self.df["sigma_V"])],
            "$\\frac{\\delta V_i}{V_i}$": [f"{(sigma_V / V):.3f}" for V, sigma
_V in zip(self.df["V"], self.df["sigma_V"])]
        })
    
    def export_latex_table(self, filename=None):
        """
        Generate LaTeX code for the formatted table.

        Parameters:
        filename : str or None
            - If None: prints the LaTeX table directly to the screen.
            - If str: saves the LaTeX table to the specified .tex file.

        Returns:
        str : The LaTeX table code.
        """
        table = self.format_table()
        latex_code = table.to_latex(escape=False, index=False)

        if filename:
            with open(filename, "w") as f:
                f.write(latex_code)
            print(f"LaTeX table exported to {filename}")
        else:
            print(latex_code)

        return latex_code

\end{lstlisting}

An example of typical usage is presented below. Just use a Jupyter notebook 
and import the Python class. First, save the Python class in a Python file (
.py). You can leave the Python notebook and the class file in the same 
folder, so you do not need to create path environment variables for the files.

\begin{lstlisting}[language=Python, caption=Example usage of WaterDensity 
class in Jupyter Notebook]
from class_water_density import WaterDensity

# Instantiate the class with default parameters
wd = WaterDensity()

# Default values
print(wd.M0, wd.V0, wd.rho)  # 300 100 1

# Redefine them manually:
wd.M0 = 200
wd.V0 = 100
wd.rho = 0.997

# Step 1: Generate fake experimental data
df = wd.gen_fake_data(
    n_points=10,  # number of data points
    V_min=110,    # minimum volume
    V_max=200,    # maximum volume
    err_M=0.02,  # relative error in mass
    err_V=0.02,   # relative error in volume
    noise=2       # random noise
)

# Step 2: Perform least squares linear regression on the generated data
tb1, fit_results = wd.calculate_fit()

# Step 3: Format the data table for LaTeX-like reporting
tb2 = wd.format_table()

# Step 4: Plot the regression line along with data and uncertainty bands
wd.plot_regression()

# Step 5: Export LaTeX table to file
table_tex = wd.export_latex_table("mass_volume_table.tex")

# print the LaTeX code
print(table_tex)
\end{lstlisting}

The reader may also refer to the following GitHub repo for more Python 
material related to this article. More usage examples will be updated, and a 
Jupyter Notebook with the example can be downloaded for study purposes. If 
the repository code is used in any publication, please refer to it via a link.

\href{https://github.com/osvaldosantospereira/water\_density\_physexp/tree/main}
{https://github.com/osvaldosantospereira/water\_density\_physexp/tree/main}

\section{Conclusion}

This study analyzed an undergraduate physics laboratory experiment designed 
to determine the density of water using fundamental measurement techniques 
and regression analysis. The experimental setup, which includes a precision 
scale, a graduated container filled with water, and a suspended metal rod, 
allows students to develop critical skills in experimental physics. 
Throughout the experiment, students are challenged to derive theoretical 
models that link physical observable variables, such as mass and volume, that 
can be experimentally measured, to the physical quantity of interest—water 
density — via a mathematical formula.

One of the main difficulties students may encounter is understanding the 
process of model linearization to achieve a linear regression via Ordinary 
Least Squares methods. Additionally, simple but often overlooked physical 
phenomena, such as frictional forces between the metal rod and the 
container's surface, can introduce systematic errors, affecting the results. 
Furthermore, students may often face conceptual challenges in selecting the 
appropriate model to analyze mass as a function of volume ($M \times V$), or 
volume as a function of mass ($V \times M$). This choice directly influences 
the error propagation and the reliability of the final water density 
calculation. 

This article addresses some of the challenges that students may face by 
providing theoretical and computational guidance to gain deeper insight into 
the physical interpretations of their experimental results. It presents key 
elements of data analysis similar to what would be expected in a lab exam or 
documentation. Python scripts are provided to fit the linear model and 
visualize the experimental data, reinforcing the importance of integrating 
computational tools into experimental physics education.

This work highlights the necessity of pedagogical approaches that bridge 
theoretical concepts with hands-on experimental work and computational tools, 
ultimately fostering a more robust understanding of data analysis and 
physical modeling in undergraduate courses of physical science and 
engineering curricula.

\begin{appendix}
\section{Ordinary Least Squares}

Given a dataset of $N$ data points $(x_i, y_i)$, the objective is to 
determine a linear model
\be
y_i = a x_i + b
\ee
where $a$ is the slope, and $b$ is the intercept, which are the free 
parameters of this model. For this, we will derive the closed formulas using 
the Least Squares Method, which minimizes the sum of squared residuals $
\varepsilon_i$
\be
\varepsilon_i = y_i - a x_i - b\,,
\ee
So, we determine the function 
\be
S(a,b) = \sum_{i=1}^{N} (y_i - ax_i - b)^2 \,.
\ee
To find $a$ and $b$, we must optimize the function $S(a,b)$ concerning its 
parameters, computing the partial derivatives of $S$ and setting them to zero.
\be
\frac{\partial S}{\partial a} = \sum_{i=1}^{N} 2(y_i - ax_i - b)(-x_i) = 0\,,
\ee
which results in the following expression
\be
\sum_{i=1}^{N} x_i y_i = a \sum_{i=1}^{N} x_i^2 + b \sum_{i=1}^{N}x_i\,, 
\label{eq1}
\ee
now performing the partial derivatives concerning $b$
\be
\frac{\partial S}{\partial b} = \sum_{i=1}^{N} 2(y_i - ax_i - b)(-1) = 0.
\ee
The above expression results in the following
\be
\sum y_i = a \sum x_i + N b \,. \label{eq2}
\ee
For simplicity, we can use the following notation
\be
S_{xy} = S_{yx} = \sum_{i=1}^{N} x_i y_i\,,
\ee
\be
S_{xx} = \sum_{i=1}^{N} x_i^2\,, 
\ee
\be
S_x = \sum_{i=1}^{N} x_i\,,
\ee
\be
S_y = \sum_{i=1}^{N} y_i
\ee
So equations Eq.\,\eqref{eq1} and Eq.\,\eqref{eq2} can be put in the form
\begin{align}
S_{xy} &= a S_{xx} + b S_x \\
S_y &= a S_x + N b
\end{align}
Or in matrix form
\be
\begin{pmatrix}
S_{xx} & S_x \\
S_x & N
\end{pmatrix}
\begin{pmatrix}
a \\ b
\end{pmatrix} = 
\begin{pmatrix}
S_{xy} \\ S_{y}
\end{pmatrix}
\ee
Solving for the system of equations given by Eq.\,\eqref{eq1} and 
Eq.\,\eqref{eq2} for $a$ and $b$ results in the following expression for the 
estimator $\hat a$ (slope)
\be
\hat{a} = \frac{N S_{xy} - S_x S_y}{N S_{xx} - S_x^2} = \frac{\sum_{i=1}^N (x_
i - \hat{x})(y_i - \hat{y})}{\sum_{i=1}^N(x_i - \hat{x})^2}. \label{expa}
\ee
Remembering that from the linear model
\be
b = \hat{y} - a \hat{x} \label{expb}
\ee
where $\hat{y}$ and $\hat{x}$ are the average estimators given by
\be
\hat{y} = \frac{1}{N}\sum_{i=1}^N y_i = \frac{S_y}{N} \,, \label{avy}
\ee
\be
\hat{x} = \frac{1}{N}\sum_{i=1}^N x_i = \frac{S_x}{N}\,, \label{avx}
\ee
Substituting Eq.\,\eqref{avy}, Eq.\,\eqref{avx} and Eq.\,\eqref{expa} into Eq.
\,\eqref{expb} results in
\be
\hat{b} = \frac{S_y S_{xx} - S_{xy}S_x}{NS_{xx} - S_x^2} = \hat{y} - \frac{
\sum_{i=1}^N (x_i - \hat{x})(y_i - \hat{y}) }{\sum_{i=1}^N(x_i - \hat{x})^2}
\hat{x}
\ee
Notice that
\begin{align}
\sum_{i=1}^N(x_i - \hat{x})^2 &= \sum_{i=1}^N x_i^2  - 2 \hat{x} \sum_{i=1}^N 
x_i + \sum_{i=1}^N \hat{x}^2 \nonumber \\ 
&= \sum_{i=1}^N x_i^2 - 2N \hat{x}^2 + N \hat{x}^2 \nonumber \\
&= \sum_{i=1}^N x_i^2 - N \hat{x} \nonumber \\
&= \sum_{i=1}^N x_i^2 - \frac{1}{N}\sum_{i=1}^N x_i
\end{align}

\section{Error in Estimator a}

A Step-by-Step Derivation of Variance and Standard Error of $\hat{b}$. To 
fully understand the derivation of the Variance and standard error of $ \hat{b
} $, we need to go deeper into the mathematics. Recall the Least Squares 
Estimate for $\hat{b}$. The least squares estimate of the slope in a simple 
linear regression model is given by
\be
\hat{a} = \frac{\sum (x_i - \bar{x}) (y_i - \bar{y})}{\sum (x_i - \bar{x})^2}
\,,
\ee
where $\bar{x} = \frac{1}{n} \sum x_i$ is the mean of $x$, and $\bar{y} = 
\frac{1}{n} \sum y_i$ is the mean of $y$. This formula tells us that $ \hat{b}
 $ is a linear function of $y_i$, which allows us to compute its Variance. 
Express $\hat{a}$ in Terms of the Error terms. The regression model assumes:
\be
y_i = a + b x_i + \varepsilon_i
\ee
where $\varepsilon_i$ are independent, normally distributed errors with mean 
zero and Variance $\sigma^2$
\be
E[\varepsilon_i] = 0, \quad \text{Var}(\varepsilon_i) = \sigma^2.
\ee
Substituting this into our equation for $\hat{a}$, leads to
\be
\hat{a} = \frac{\sum (x_i - \bar{x}) (b + a x_i + \varepsilon_i - \bar{y})}{
\sum (x_i - \bar{x})^2}.
\ee
Expanding $ \bar{y} = a + b \bar{x} + \bar{\varepsilon} $:
\be
\hat{a} = \frac{\sum (x_i - \bar{x}) (b + a x_i + \varepsilon_i - (b + a \bar{
x} + \bar{\varepsilon}))}{\sum (x_i - \bar{x})^2}.
\ee
Since $ \sum (x_i - \bar{x}) \bar{\varepsilon} = 0 $, simplifying gives
\be
\hat{a} = a + \frac{\sum (x_i - \bar{x}) \varepsilon_i}{\sum (x_i - \bar{x})^2
}.
\ee
Compute the Variance of $\hat{a}$ by taking the Variance of both sides
\be
\text{Var}(\hat{a}) = \text{Var} \left( \frac{\sum (x_i - \bar{x}) \varepsilon
_i}{\sum (x_i - \bar{x})^2} \right).
\ee
Since the errors $\varepsilon_i$ are independent and have variance $\sigma^2$
\be
\text{Var} \left( \sum (x_i - \bar{x}) \varepsilon_i \right) = \sum (x_i - 
\bar{x})^2 \text{Var}(\varepsilon_i) = \sigma^2 \sum (x_i - \bar{x})^2.
\ee
Since variance scales by $1/k^2$ when dividing by a constant $k$, we get:
\be
\text{Var}(\hat{a}) = \frac{\sigma^2 \sum (x_i - \bar{x})^2}{\left(\sum (x_i -
 \bar{x})^2\right)^2} = \frac{\sigma^2}{\sum (x_i - \bar{x})^2}.
\ee
Thus, the error on the estimator $\hat{a}$ is given by
\be
\sigma_{\hat{a}} = \frac{\sigma}{\sqrt{\sum (x_i - \bar{x})^2}}.
\ee

\section{Error in Estimator b}

To derive the Variance and standard error of the estimator $\hat{b}$ in the 
regression equation
\be
y = ax + b + \varepsilon,
\ee
we start with its least squares estimate:
\be
\hat{b} = \bar{y} - \hat{a} \bar{x}.
\ee
Substituting $\hat{a}$,
\be
\hat{b} = \bar{y} - \frac{\sum (x_i - \bar{x}) (y_i - \bar{y})}{\sum (x_i - 
\bar{x})^2} \bar{x}.
\ee
Expressing in terms of the true model, 
\be
\bar{y} = a \bar{x} + b + \bar{\varepsilon}.
\ee
Thus,
\be
\hat{b} = a \bar{x} + b + \bar{\varepsilon} - \frac{\sum (x_i - \bar{x}) 
\varepsilon_i}{\sum (x_i - \bar{x})^2} \bar{x}.
\ee
Simplifying,
\be
\hat{b} = b + \bar{\varepsilon} - \frac{\sum (x_i - \bar{x}) \varepsilon_i}{
\sum (x_i - \bar{x})^2} \bar{x}.
\ee
Taking variances,
\be
\text{Var}(\hat{b}) = \text{Var} \left( \bar{\varepsilon} - \frac{\sum (x_i - 
\bar{x}) \varepsilon_i}{\sum (x_i - \bar{x})^2} \bar{x} \right).
\ee
Since 
\be
\text{Var}(\bar{\varepsilon}) = \frac{\sigma^2}{n}
\ee 
and
\be
\text{Var} \left( \frac{\sum (x_i - \bar{x}) \varepsilon_i}{\sum (x_i - \bar{x
})^2} \right) = \frac{\sigma^2}{\sum (x_i - \bar{x})^2},
\ee
we use the property 
\be
\text{Var}(A + B) = \text{Var}(A) + \text{Var}(B) + 2\text{Cov}(A, B) 
\ee
and the known covariance result
\be
\text{Cov} \left(\bar{\varepsilon}, \frac{\sum (x_i - \bar{x}) \varepsilon_i}{
\sum (x_i - \bar{x})^2} \right) = -\frac{\sigma^2 \bar{x}}{\sum (x_i - \bar{x}
)^2}.
\ee
Thus,
\be
\text{Var}(\hat{b}) = \frac{\sigma^2}{n} + \frac{\sigma^2 \bar{x}^2}{\sum (x_
i - \bar{x})^2}.
\ee
Taking the square root,
\be
\sigma_{\hat{b}} = \sqrt{\sigma^2 \left( \frac{1}{n} + \frac{\bar{x}^2}{\sum (
x_i - \bar{x})^2} \right)}.
\ee
Simplifying even further results in the following expression
\be
\sigma_{\hat{b}} = \sqrt{\frac{\sum_{i=1}^N x_i^2}{\sum (x_i - \bar{x})^2}} 
\sigma.
\ee

\end{appendix}

\end{document}